\documentclass[aps,prb,twocolumn,floatfix,unsortedaddress,citeautoscript]{revtex4}

\usepackage{graphicx}
\usepackage{amsmath}
\graphicspath{{./}{./EPS/}{./Figures/}}

\begin{document}

\title{Persistent current in ballistic mesoscopic rings with
Rashba spin-orbit coupling}

\author{Janine Splettstoesser}
\affiliation{Institut f\"ur Theoretische Festk\"orperphysik,
Universit\"at Karlsruhe, D-76128 Karlsruhe, Germany}

\author{Michele Governale}
\affiliation{Institut f\"ur Theoretische Festk\"orperphysik,
Universit\"at Karlsruhe, D-76128 Karlsruhe, Germany}
\affiliation{NEST-INFM \& Scuola Normale Superiore, I-56126 Pisa,
Italy}

\author{Ulrich Z\"ulicke}
\affiliation{Institut f\"ur Theoretische Festk\"orperphysik,
Universit\"at Karlsruhe, D-76128 Karlsruhe, Germany}

\date{\today}

\begin{abstract}
The presence of spin--orbit coupling affects the spontaneously
flowing persistent currents in mesoscopic conducting rings. Here
we analyze their dependence on magnetic flux with emphasis on
identifying possibilities to prove the presence and extract the
strength of Rashba spin splitting in low--dimensional systems.
Effects of disorder and mixing between quasi--onedimensional ring
subbands are considered. The spin-orbit coupling strength can be
inferred from the values of flux where sign changes occur in the
persistent charge current. As an important consequence of the
presence of spin splitting, we identify a nontrivial persistent
spin current that is not simply proportional to the
charge current. The different flux dependences of persistent
charge and spin currents are a {\em unique\/} signature of
spin--orbit coupling affecting the electronic structure of the
ring.
\end{abstract}


 
\maketitle

\section{Introduction}

The interplay between spin-orbit (SO) coupling and quantum
confinement in semiconductor heterostructures has recently
attracted great interest. It provides a useful tool to manipulate
the spin degree of freedom of electrons  by coupling to their
orbital motion, and vice versa. As a result, spin-orbit coupling
has become one of the key ingredients for phase-coherent
spintronics applications\cite{lossbook,wolf}. Various sources of
broken inversion symmetry give rise to intrinsic (zero--field)
spin splitting in semiconductor heterostructures\cite{lommer}.
We focus here on the one induced by structural inversion
asymmetry, i.e., the Rashba effect\cite{rashba}. It is typically
important in small-gap zinc--blende--type semiconductors and can
be tuned by external gate voltages\cite{nitta,schaepers,grundler}.

Many proposals have been put forward recently for devices based on
spin-dependent transport effects due to the Rashba SO coupling in
low-dimensional systems\cite{devices}. To explore possibilities
for their realization, it is desirable to have a reliable way to
determine experimentally the strength $\alpha$ of the Rashba SO
coupling. Transport experiments have been performed in
two-dimensional (2D) electron systems, and $\alpha$ was extracted
from beating patterns in the Shubnikov-de~Haas
oscillations\cite{nitta,schaepers,grundler} as well as the SO
relaxation time obtained from weak--antilocalization behavior in
the resistance\cite{weakal}. The only previous experimental
studies of SO coupling in {\em quasi--1D\/} systems have measured
transport through mesoscopic rings\cite{morpurgo,shayegan}.
Beating patterns in the Aharonov-Bohm (AB) oscillations of the
ring's conductance are expected to arise from quantum
phases\cite{loss:prl:90,aronov,qiansu,diegoprb} induced by the
presence of SO coupling. 

In practice, it turns out\cite{navalexp}, however, that the
signature of the Rashba effect in AB oscillations can be masked by
features arising due to the ring's nonideal coupling to external
leads. As an alternative, we explore here the possibility to
obtain a direct measure of the Rashba SO coupling strength from
the persistent current\cite{1dxstal,cheung} induced by a magnetic
flux perpendicular to the ring. This approach would have the
advantage of circumventing entirely any problems arising from
contacting the ring.

There is a vast literature of theoretical\cite{1dxstal,cheung,
loss_goldbart,bouchiat,montambaux,meir_gefen,entin-wohlmann,
mathur,chinese} and experimental\cite{levi,chandra,mailly} studies
on persistent currents. From the theoretical point of view, the
effect of SO coupling on the Fourier transform of observables has
been addressed in Refs.~\onlinecite{meir_gefen,entin-wohlmann,
mathur}. Measurements of the persistent charge current have been
performed both in an ensemble of metallic rings\cite{levi} and on
single isolated rings realized in nanostructured 2D electron
systems\cite{chandra,mailly}. So far, persistent currents have not
yet been studied in rings where the Rashba effect is likely to be
important. From our study, we find features in the flux dependence
of the persistent charge current that allow for a direct
quantitative determination of the Rashba SO coupling strength. We
discuss how averaging over rings with different numbers of
particles and mixing between different 1D subbands affects these
features. An unambiguous signature of SO coupling is obtained from
a comparison of the persistent {\em spin\/} current with the
persistent {\em charge\/} current. In the absence of SO coupling, the
persistent spin current is finite only for an odd number of
particles in the ring and is proportional to the persistent charge
current. With SO coupling, the persistent spin current is finite
also for an even electron number. For an odd number of electrons
in the ring, the persistent spin current is sizeable only for
small values of the SO coupling strength. The flux dependence of
the persistent spin current is generally strikingly different from
that of the charge current. Observability of the persistent spin
current by its induced electric field\cite{meier-loss,schuetz,
hongguo,flavio} should enable the unambiguous identification of SO
effects in low--dimensional mesoscopic rings.

The paper is organized as follows. In Section~\ref{secmodel}, we
write down and discuss the model Hamiltonian used to describe the
ring. Electronic properties and persistent currents of a purely
1D ring are computed in the following Sec.~\ref{sec1d}.
Section~\ref{sec2d} is devoted to the effect of higher radial
subbands. Conclusions are presented in Sec.~\ref{conclusions}. 

\section{Model of a mesoscopic ring with Rashba spin--orbit
coupling} \label{secmodel}

For completeness and to introduce notation used later in our work,
we outline here briefly the derivation of the Hamiltonian
describing the motion of an electron in a realistic quasi--1D
ring\cite{meijer}. We consider 2D electrons in the $xy$ plane that
are further confined to move in a ring by a radial potential
$V_{\text{c}}(r)$. The electrons are subject to the Rashba SO
coupling, which reads
\begin{equation}  
\label{hso}
H_{\text{so}}=\frac{\alpha}{\hbar}\left(\sigma_{x}\,(\vec{p}-e
\vec{A})_{y}-\sigma_{y}\,(\vec{p}-e\vec{A})_{x}\right).  
\end{equation}
Here $\vec{A}$ is the vector potential of an external magnetic
field applied in the $z$ direction. The coupling strength $\alpha$
defines the spin-precession length $l_{\text{so}}=\pi\hbar^2/(m
\alpha)$. The full single-electron Hamiltonian reads 
\begin{equation}
H= \frac{(\vec{p}-e \vec{A})_{x}^{2}+
(\vec{p}-e \vec{A})_{y}^{2}}{2m}
+V_{\text{c}}(r)+H_{\text{so}}+\hbar\omega_{\text{z}}\sigma_{z},
\end{equation}
where the Zeeman splitting from the external magnetic field is
included as the last term. Due to the circular symmetry of the
problem, it is natural to rewrite the Hamiltonian in polar
coordinates:\cite{meijer}
\begin{eqnarray}
\nonumber
H&=&-\frac{\hbar^2}{2m}\left[\frac{\partial^2}{\partial r^2}+
\frac{1}{r}\frac{\partial}{\partial r}-\frac{1}{r^2}\left(i 
\frac{\partial}{\partial \varphi} +\frac{\Phi}{\Phi_0}\right)^2
\right]+V_{\text{c}}(r)
\\
\label{polarh} 
& & -\frac{\alpha}{r} \sigma_r 
\left( i \frac{\partial}{\partial{\varphi}}+\frac{\Phi}{\Phi_0}
\right)+i \alpha \sigma_\varphi \frac{\partial}{\partial r}+ 
\hbar\omega_{\text{z}}\sigma_{z}, 
\end{eqnarray} 
where $\Phi$ is the magnetic flux threading the ring, $\Phi_0$ the
flux quantum, $\sigma_r=\cos\varphi\,\sigma_x+\sin\varphi\,
\sigma_y$ and $\sigma_\varphi=-\sin\varphi\,\sigma_x+\cos\varphi\,
\sigma_y$. In the case of a thin ring, i.e., when the radius $a$
of the ring is much larger than the radial width of the wave
function, it is convenient to project the Hamiltonian on the
eigenstates of $H_0=-\frac{\hbar^2}{2m}\left[\frac{\partial^2}
{\partial r^2}+\frac{1}{r}\frac{\partial}{\partial r}\right]+
V_{\text{c}}(r)$. To be specific, we use a parabolic radial
confining potential,
\begin{equation}
V_{\text{c}}(r)=\frac{1}{2} m {\omega^2}(r-a)^2 \quad ,
\end{equation}
for which the radial width of the wave function is given by
$l_\omega=\sqrt{\hbar /m \omega}$. In the following, we assume
$l_\omega/a \ll 1$ and neglect contributions of order $l_\omega/a
$. In this limit, $H_0$ reduces to
\begin{equation}
\label{h0}
H_0=-\frac{\hbar^2}{2m}
\left[\frac{\partial^2}{\partial r^2}\right]+\frac{1}{2} m
{\omega^2}(r-a)^2 \quad .
\end{equation}
We now calculate matrix elements of the Hamiltonian
Eq.~(\ref{polarh}) in the basis of eigenfunctions of
Eq.~(\ref{h0}) that correspond to quasi--1D radial subbands,
labeled here by the quantum number $n$. The diagonal matrix
elements are given by
\begin{eqnarray}
\nonumber
H_{n,n}&=&\frac{\hbar^2}{2m a^2}\left(i 
\frac{\partial}{\partial \varphi} +\frac{\Phi}{\Phi_0}\right)^2
-\frac{\alpha}{a} \sigma_r 
\left( i \frac{\partial}{\partial{\varphi}}+\frac{\Phi}{\Phi_0}
\right)\\
\label{hdiagonal}
& &-i \frac{\alpha}{2a} \sigma_\varphi 
 +\hbar\omega_{\text{z}}\sigma_{z}+\hbar \omega (n+\frac{1}{2})
\quad .  
\end{eqnarray}
The only nonvanishing offdiagonal matrix elements are those
coupling adjacent radial subbands:  
\begin{equation}
\label{hmix}
H_{n,n+1}=H_{n+1,n}^{\dagger}= 
i \sigma_{\varphi}\sqrt{\frac{n+1}{2}}\frac{\alpha}{l_\omega}.  
\end{equation}

\section{Properties of ideal 1D rings}
\label{sec1d}

The ideal 1D limit for a mesoscopic ring is realized when only the
lowest radial subband is occupied by electrons and all relevant
energy scales as, e.g., temperature, voltage, and disorder
broadening are small enough such that interband excitations can be
neglected. In the following Section, we focus on this situation
that can be realized in recently fabricated ring
structures\cite{lorke,heinzel,haug}.

\subsection{Energy spectrum of 1D ring with impurity}
\label{1dspectrum}

\begin{figure}[t]
\includegraphics[width=3.in]{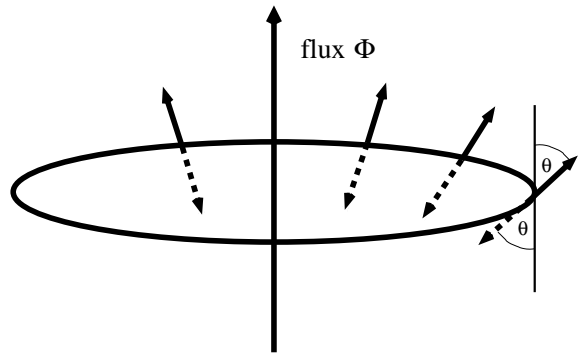}
\caption{Schematic illustration of the spin texture exhibited by 
the eigenstates of the ideal one-dimensional ring.
\label{texture}}
\end{figure}  
Straightforward algebra yields the eigenenergies of $H_{0,0}$
which are usually labeled by an integer number $q$:
\begin{eqnarray}
\nonumber
E_{q,\pm}&=&\hbar \omega_{\text{a}} \left(q-\frac{\Phi}{\Phi_0}+
\frac{1}{2}\mp\frac{1}{2 \cos\theta_q}\right)^2\\
\label{eigenexact}
& &+\frac{\hbar \omega_{\text{a}}}{4} 
\left(1-\frac{1}{\cos^2\theta_q}\right)
\pm \frac{\hbar\omega_{\text{z}}}{\cos\theta_q} .
\end{eqnarray}
Here we have introduced the frequency $\omega_{\text{a}}=\hbar/(2
m a^2)$ and omitted the constant energy shift of the radial
subband bottom. The eigenvectors corresponding to the
eigenenergies given in Eq.~(\ref{eigenexact}) are 
\begin{equation}
\Psi_{q,\pm}=e^{i(q+\frac{1}{2})\varphi} \chi_{q,\pm},
\end{equation}
with the spinors
\begin{subequations}
\label{vectors}
\begin{eqnarray}
\chi_{q,+}&=&
\left(\begin{array}{c}\cos(\frac{\theta_q}{2})e^{-i\frac{1}{2}
\varphi}
\\ \sin(\frac{\theta_q}{2})e^{i\frac{1}{2}\varphi}\end{array}
\right)\quad , \\ 
\chi_{q,-} &=&\left(\begin{array}{c}-
\sin(\frac{\theta_q}{2})e^{-i\frac{1}{2}\varphi} \\ 
\cos(\frac{\theta_q}{2})e^{i\frac{1}{2}\varphi}\end{array}\right)
\quad .
\end{eqnarray}
\end{subequations}
The angle $\theta_q$ is given by\cite{aronov} 
\begin{equation}
\label{tantheta}
\tan(\theta_q)=-\frac{\frac{\alpha}{a}(q-\frac{\Phi}{\Phi_{0}}+
\frac{1}{2})}{\hbar\omega_{\text{a}}(q-\frac{\Phi}{\Phi_{0}}+
\frac{1}{2}) -\hbar\omega_{\text{z}}}.
\end{equation} 
\begin{figure}[t]
\includegraphics[width=2.9in]{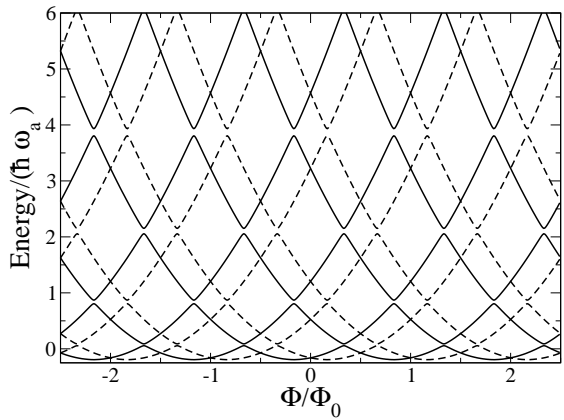}
\caption{Single-particle energy spectrum of an ideal 1D ring with
a model delta--barrier impurity. Parameters are $\cos\theta=2/5$,
and $A=0.1$. Energy levels for states corresponding to spin-up
(solid line) and spin-down (dashed line) in the local-spin-frame
basis are shifted, in flux direction, by $1/\cos\theta$. 
\label{spectrum}}
\end{figure} 
The spinors $\chi_{q,\pm}$ are the eigenstates of the operator 
\begin{equation}\label{locspinmat}
\sigma_{\theta_q}=\sigma_z\,\cos\theta_q+\sigma_r\,\sin\theta_q
\quad ,
\end{equation}
and constitute a basis in spin space with space-dependent
quantization direction, as shown in Fig.~\ref{texture}. We will
refer to this $\varphi$-dependent spin basis as the {\em local spin
frame}. $\theta_q$ is the angle between the local quantization
axis and the direction perpendicular to the ring ($z$ axis). The
tilt angle described by Eq.~(\ref{tantheta}) becomes independent
of the quantum number $q$ when the Zeeman energy is negligible,
i.e., when $\left|\hbar\omega_{\text{a}}(q-\frac{\Phi}{\Phi_{0}}+
\frac{1}{2})\right|\gg\hbar\omega_{\text{z}}$. For typical
realizations of mesoscopic rings with many electrons present,
states contributing importantly to the persistent current fulfill
this requirement. Therefore, in the following, we focus
exclusively on the limit where Zeeman splitting vanishes and
$\theta_q\to\theta=\lim_{\omega_{\mathrm z}\to 0}\theta_q$.
Then all eigenstates have the same local spin frame, to which we
can transform using the SU(2) matrix
\begin{equation}
{\mathcal U}=\left(\begin{array}{cc} e^{-i\varphi/2}\cos\frac
{\theta}{2} & -e^{-i\varphi/2}\sin\frac{\theta}{2} \\
e^{i\varphi/2}\sin\frac{\theta}{2} & e^{i\varphi/2}\cos\frac
{\theta}{2} \end{array}\right)\quad .
\end{equation}
This yields $H_{\text{1D}}\equiv
{\mathcal U}^\dagger\,\left(H_{0,0}-\hbar\omega/2\right)_{\omega_z
=0}\,{\mathcal U}$ where
\begin{eqnarray}
\nonumber
\label{h1d}
H_{\text{1D}} &=&\hbar \omega_{\text{a}} \left(-i \frac{\partial}
{\partial\varphi}-\frac{\Phi}{\Phi_0}-
\frac{1}{2 \cos\theta}\sigma_z\right)^2\\
& &+\frac{\hbar \omega_{\text{a}}}{4}
(1-\frac{1}{\cos^2\theta}) \quad .
\end{eqnarray} 
Here $\cos \theta$ parameterizes the strength of the SO coupling.
The eigenstates in the local spin frame are simply $e^{i(q+
\frac{1}{2})\varphi}|\pm\rangle$, where $|\pm\rangle$ denote the
eigenspinors of $\sigma_z$, and the eigenenergies are given by
Eq.~(\ref{eigenexact}) with $\theta_q\to\theta$ and
$\omega_{\text{z}}=0$. Note that the orbital part of the
eigenstates obeys antiperiodic boundary conditions to compensate
for the antiperiodicity of the spinors of Eq.~(\ref{vectors}).

To discuss the effect of a nonmagnetic impurity, we exploit the
formal analogy between a ring with an impurity and a 1D periodic
potential\cite{1dxstal}. The latter is described by a
Kronig--Penney model\cite{ashcroft}, with the magnetic flux
playing the role of the quasimomentum of the 1D crystal.  
The impurity is modeled by its energy--dependent transmission
amplitude $t=|t|\exp{(i \delta)}$. The energy spectrum for the
electrons with spin $|\pm\rangle$ can now be obtained by solving
the transcendental secular equation
\begin{equation}
\label{secular0}
|t|\cos\left[2\pi\left(\frac{\Phi}{\Phi_{\mathrm{0}}}\pm\frac{1}
{2\cos(\theta)}\right)\right] =  -\cos\left(2\pi\kappa_\pm+\delta
\right),
\end{equation}
complemented by the relation  
\begin{equation}
\label{enkappa}
E_\pm=\hbar \omega_{\text{a}} \left[\kappa_\pm^2+\frac{1}{4}
(1-\frac{1}{\cos^2\theta})\right].
\end{equation}
In general, the secular equation (\ref{secular0}) cannot be solved
ana\-lytically for arbitrary transmission function $t$. To
simplify the problem, we will now assume that the impurity is a
delta-function barrier $V_0 \delta(\varphi)$. The transmission
coefficient for a state $\exp(i \kappa \varphi)|\pm\rangle$ is $t=
2 \kappa/[2\kappa + i V_0/(\hbar\omega_{\text{a}})]$. For states
close to the Fermi level, Eq.~(\ref{secular0}) can be written as
\begin{equation}
\label{secular}
\cos\left(2\pi \frac{\Phi_{\pm}}{\Phi_0}\right) = 
\cos\left(2\pi\kappa_{\pm}\right)+ \text{sign}(\kappa_{\pm}) 
A \sin \left(2 \pi\kappa_{\pm}\right),
\end{equation}
with a constant $A=V_0/(\hbar \omega_{\text{a}} \mathcal{N})$,
where $\mathcal{N}$ is the total number of electrons. We also
defined the effective fluxes
\begin{equation}
\label{flux}
\Phi_{\pm}=\Phi+\Phi_0 \left(\frac{1}{2}\pm\frac{1}
{2\cos(\theta)}\right).
\end{equation} 
Equation~(\ref{secular}) with constant $A$ would be exact for a
barrier with energy-independent transmission amplitude $t=[1-i\,
A\,\text{sign}(\kappa)]/(A^2+1)$. The approximated secular
equation (\ref{secular}) has the solution
\begin{widetext}
\begin{equation}
\label{solution}
\kappa_{q,\pm} = q + \frac{1}{2\pi}
\text{arcos}\left[\frac{\cos(2\pi \frac{\Phi_{\pm}}{\Phi_0})-
\text{sign}(q)
\sqrt{A^2\left(\sin^2 (2\pi \frac{\Phi_{\pm}}{\Phi_0})+
A^2\right)}}{1+A^2}
\right] \quad .
\end{equation}
\end{widetext}
Equation~(\ref{solution}) together with Eq.~(\ref{enkappa}) yields
the single-particle energy spectrum for the ring with an idealized
impurity. Note that, in the representation of the local spin
frame, the impurity problem maps to that of electrons without SO
coupling but with an effective spin--dependent
flux\cite{meir_gefen,entin-wohlmann} given by Eq.~(\ref{flux}).
This is illustrated in an example spectrum shown in
Fig.~\ref{spectrum}.

\subsection{Persistent charge currents}

Having calculated the single-particle electronic properties of 
the ring, we proceed to evaluate the persistent charge current. At
zero temperature, it is given by\cite{1dxstal}
\begin{equation}
\label{persistent}
I=-\frac{\partial E_{\text{gs}}}{\partial \Phi}=
-\sum_{i \in \text{occupied}} \frac{\partial E_i} {\partial \Phi},
\end{equation}
where $E_{\text{gs}}$ is the ground state energy, and $E_i$ are 
the single particle eigenenergies. Here $i$ stands for a set of
quantum numbers used to label corresponding eigenstates, including
here the spin projection in the local spin frame. The second
equality in Eq.~(\ref{persistent}) is valid only in the absence of
electron-electron interactions, which we neglect here. The
zero-temperature formula applies when the thermal energy
$k_{\text{B}} T$ is smaller than the energy difference between the
last occupied state and the first unoccupied one. In the
following, we will always consider the number $\mathcal{N}$ of
electrons in the ring to be fixed, i.e., work in the canonical
ensemble. This is the relevant situation for an isolated ring.

\begin{figure}
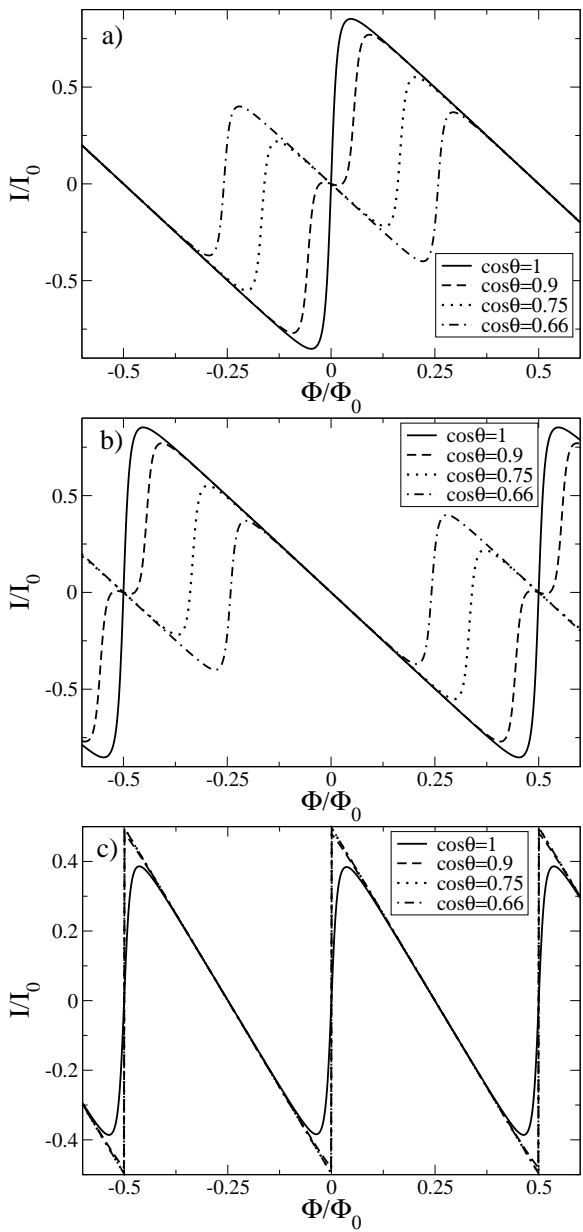

\includegraphics[width=3.in]{fig3a.eps}\\
\includegraphics[width=3.in]{fig3b.eps}\\
\includegraphics[width=3.in]{fig3c.eps}
\caption{Persistent charge current vs.\ magnetic flux for a set of
values for the spin-orbit coupling strength. The total number of
electrons is set to $4 N$ in panel a), to $4N+2$ in panel b), and
to $2 N+1$ in panel c) in the regime of large-enough $N$ such that
the persistent current is universal. A dimensionless barrier
strength of $A=0.1$ was assumed. The persistent current is
measured in units of $I_0=\hbar \omega_{\text{a}} \mathcal{N}/
\Phi_0$.
\label{persistent1}}
\end{figure}

\begin{figure}
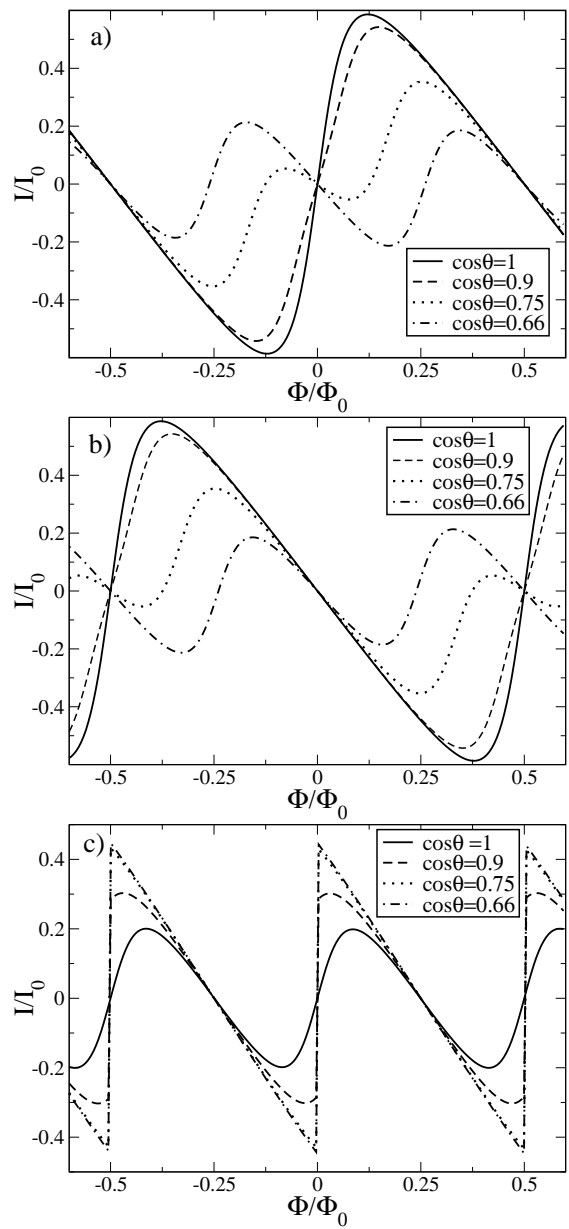

\includegraphics[width=2.9in]{fig4a.eps}\\
\includegraphics[width=2.9in]{fig4b.eps}\\
\includegraphics[width=2.9in]{fig4c.eps}
\caption{Same as Fig.~\ref{persistent1} but with different
impurity parameter $A=0.5$. Note the remaining sharpness of
jumps in the case of odd electron number even at this rather large
value of $A$.
\label{persistent2}}
\end{figure}

For spinful electrons, the flux dependence of the persistent
charge current is distinctly different for the following
cases:\cite{loss_goldbart} i)~$\mathcal{N}=4N$, ii)~$\mathcal{N}=4
N +2$, and iii)~$\mathcal{N}=2N +1$, where $N$ denotes a
positive integer. When $\mathcal{N}$ is large enough, the
persistent charge current in units of $I_0=\hbar\omega_{\text{a}}
\mathcal{N}/\Phi_0$ has a universal behavior independent of
$\mathcal{N}$. We start discussing the weak barrier limit (small
$A$ in our model), shown in Fig.~\ref{persistent1}. In the case~i)
where $\mathcal{N}=4N$, the numbers of spin-up and spin-down
electrons (spin projection in the local spin frame!) are both
even, resulting in jumps of the persistent current at $\Phi/\Phi_0
=M+1/2\pm 1/(2 \cos \theta)$, with $M$ being integer. This is
simply the superposition of the even--number spinless--electron
persistent--current characteristics for each spin direction,
shifted in flux by $\pm 1/(2\cos\theta)$. Case~ii) corresponds to
an odd number of spin-up and spin-down electrons and exhibits
jumps of the persistent charge current at $\Phi/\Phi_0=M\pm 1/(2
\cos\theta)$, which is the analogous superposition of the
appropriately flux--shifted spinless odd-electron currents for
each spin direction. Note that the case $\mathcal{N}=4N+2$ is
obtained from the $\mathcal{N}=4N$ case simply by shifting flux by
$1/2 \Phi_0$. It is apparent that, for both cases~i) and (ii), the
minimum distance between jumps of the persistent charge current
within the periodic flux interval is a measure of $1/\cos\theta$
and, hence, of the SO coupling strength. In contrast, for
case~iii), i.e., an odd number of electrons in the ring, jumps
appear at the same values of flux ($\Phi/\Phi_0=0$ and $\pm 1/2$)
as in the absence of spin--orbit coupling. The only effect of SO
coupling turns out to be a suppression of impurity rounding for
these jumps. This can be explained quite easily. Inspection shows
that, for finite SO coupling, jumps in the persistent charge
current in the case of an odd number of electrons are due to
crossing of levels with opposite spin, while those in the case of
even electron number arise from crossings of levels having the
same spin. As a spin-independent impurity cannot couple levels
with opposite spin, only the jumps in the case of even electron
number get rounded because of impurity-induced anticrossings. For
an odd number of electrons, jumps in the persistent charge current
get broadened only by temperature. The effect of increasing
impurity (barrier) strength can be seen comparing
Fig.~\ref{persistent1} and Fig.~\ref{persistent2}, where the
persistent charge current is shown for different SO coupling
strengths, occupancy of the ring, and disorder. 
 

\begin{figure}[t]
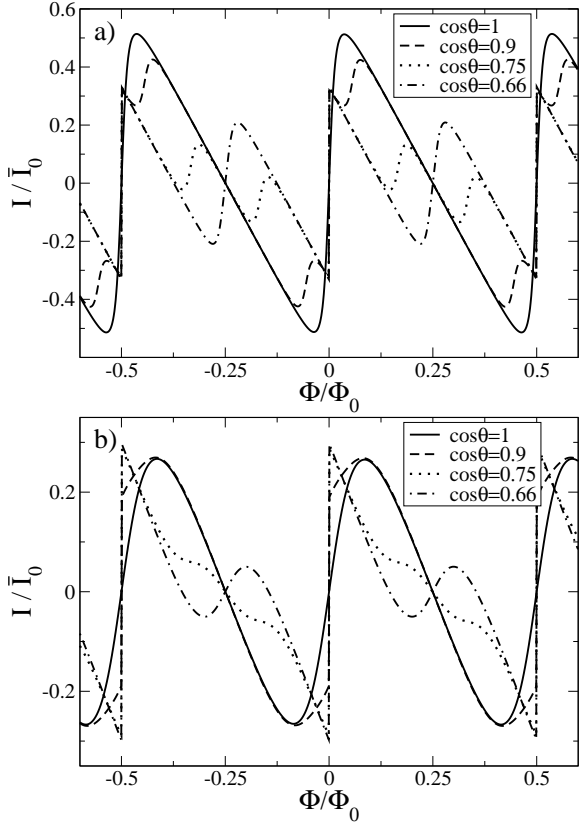

\includegraphics[width=3.in]{fig5a.eps}\\
\includegraphics[width=3.in]{fig5b.eps}\\
\caption{Average persistent charge current for an ensemble of
identical rings with different electron numbers, shown as function
of magnetic flux for  different values of the spin-orbit coupling
strength. The impurity parameter is $A=0.1$ in panel~a) and
$A=0.5$ in panel~b). The current unit is $\bar{I}_0=\hbar
\omega_{\text{a}} \bar{\mathcal{N}}/\Phi_0$, where
$\bar{\mathcal{N}}$ denotes the average number of electrons. 
\label{average}}
\end{figure}

Measurements are often performed on ensembles of many
rings\cite{levi}. The measured persistent charge current is then
an average over different occupation numbers, with even and odd
occupation occurring with the same probability. Among cases with
even electron numbers, ${\mathcal N}=4N$ and $4N+2$ would also be
equiprobable. An example of average persistent charge current is
shown in Fig.~\ref{average}. It exhibits the well--known period 
halving\cite{bouchiat,montambaux} which must occur irrespective of
the presence of SO coupling. Most importantly, however, all the
features present for the single ring and discussed above for
different occupancy are still visible. It should therefore be
possible to obtain the Rashba SO coupling strength from a
measurement of the ensemble-averaged persistent charge current. 

\subsection{Persistent spin currents} 

As electrons carry spin as well as charge, their motion gives rise
also to a spin current besides the charge current. Very often, the
difference of charge currents carried by spin-up and spin-down
electrons is identified with the spin current. While this is
appropriate in many contexts, it has to be kept in
mind\cite{loss:prl:90,flavio} that the spin current is actually a
tensor. A particular case where this fact matters is the one to be
considered here. As the electron velocity in the presence of SO
coupling turns out to be an operator in spin
space\cite{uz:prl:02}, and eigenstates for electrons of the ring
correspond to eigenspinors of a spatially
varying spin matrix [$\sigma_\theta$ as defined in
Eq.~(\ref{locspinmat})], the proper expression for the spin
current has to be derived carefully. After presenting details of
this derivation, we proceed to show results for the persistent
spin currents of electrons in a ring with Rashba SO coupling.

The operator of the $\nu$ component of spin density in real-space
representation is given by $s_\nu(\vec r) = \sigma_\nu(\vec
{r^\prime})\,\delta(\vec r - \vec{r^\prime})$, with $\sigma_\nu$
being the SU(2) spin matrix whose eigenstates form the basis for
projection of spin in $\nu$ direction. In general, this projection
direction can vary in space. The equation of motion for the
spin-density operator is given by the familiar Heisenberg form
\begin{subequations}
\begin{eqnarray}\label{spineom}
&& \frac{d}{d t} \, s_\nu(\vec r) = \frac{i}{\hbar}\,\left[ H\, ,
\, s_\nu(\vec r)\right] \quad ,\\
&& = \left(\frac{d}{dt}\,\sigma_\nu(\vec{r^\prime})\right)\,\delta
(\vec r - \vec{r^\prime}) - \vec\nabla_{\vec r} \cdot \left(
\sigma_\nu(\vec{r^\prime})\, \vec v(\vec r)\right)\quad .
\end{eqnarray}
\end{subequations}
Here $\vec\nabla_{\vec r}$ denotes the gradient operator acting on
the coordinate $\vec r$, and $\vec v(\vec r)$ is the electron
velocity operator. The latter differs from its expression $\vec
v_0$ in the absence of SO coupling by a spin--dependent
term:\cite{uz:prl:02} $\vec v=\vec v_0+\alpha(\hat z\times\vec
\sigma)/\hbar$.

Straightforward calculation for the case of {\em spatially constant\/}
$\sigma_\nu$ and vanishing Zeeman splitting yields the continuity
equation
\begin{subequations}
\begin{equation}\label{constcont}
\frac{d}{d t} \, s_\nu(\vec r) + \vec\nabla\cdot{\vec\jmath}_\nu
(\vec r) = \frac{2\alpha}{\hbar^2}\,\left(\hat\nu\times(\hat z
\times\vec\sigma)\right)\cdot\left(\vec p - e \vec A \right),
\end{equation}
with the $\nu$ component of the spin-current tensor given by
\begin{equation}\label{spincurr1}
{\vec\jmath}_\nu(\vec r) = \vec v(\vec r)\,\sigma_\nu \quad .
\end{equation}
\end{subequations}
We have used the symbols $\hat z$ and $\hat\nu$ to denote unit
vectors in $z$ and $\nu$ direction, respectively. Note that the
expression (\ref{spincurr1}) and the source term on the r.h.s.\
of Eq.~(\ref{constcont}) have been written in the usual shorthand
notation where it is understood that the real part has to be taken
in the expectation value. As an example, we fix $\nu=z$ and
consider the case of electrons moving in the lowest quasi-1D
radial ring subband. We find, after transformation into the
representation of the local spin frame, for the continuity
equation (\ref{constcont}) the simple expression
\begin{subequations}
\begin{equation}
\frac{d}{d t} \, s_z(\varphi) +\frac{1}{a}\frac{\partial}
{\partial\varphi}\,j_z^\varphi(\varphi) = 2 \omega_{\text{a}}\,
\sigma_y\left(i\frac{\partial}{\partial\varphi}+\frac{\phi}
{\phi_0} \right)\tan\theta\quad .
\end{equation}
The only nonvanising ($\varphi$) component of the spin current
turns out to be
\begin{eqnarray}
j_z^\varphi(\varphi)&=&\frac{\hbar}{m a}\left\{\left(-i\frac
{\partial}{\partial\varphi} - \frac{\phi}{\phi_0} - \frac{1}
{2\cos\theta}\,\sigma_z\right)\sigma_z\cos\theta\right.\nonumber
\\ &&\left. -\left(-i\frac{\partial}{\partial\varphi} - \frac
{\phi}{\phi_0}\right)\sigma_x\sin\theta\right\}\, .
\end{eqnarray}
\end{subequations}
Eigenstates on the ring which are labeled by quantum numbers $q$
and $\sigma$ carry a current for the $z$ projection of spin given
by
\begin{equation}
I_{z}^{(q\sigma)}=\frac{1}{2\pi a}\left\langle
j_z^\varphi(\varphi)\right\rangle_{q\sigma} = -\frac{1}{e}\frac
{\partial E_{q,\sigma}}{\partial\Phi}\,\sigma\cos\theta\quad ,
\end{equation}
which is just the charge current multiplied by the magnetization
in $z$ direction of the corresponding state\cite{lossnote}.

As an important example for the current of a spatially varying
projection of the magnetization, we consider the case of the local
spin frame, i.e., $\sigma_\nu(\vec{r^\prime})=\sigma_\theta(
\varphi)$. [See Eq.~(\ref{locspinmat}).] Additional terms arising
from derivatives of $\sigma_\theta$ w.r.t.\ polar angle $\varphi$
appear in the continuity equation for $s_\theta(\vec r)$. After
transformation into the local spin frame, it has the extremely
simple form
\begin{subequations}
\begin{equation}
\frac{d}{d t} \, s_\theta(\varphi) +\frac{1}{a}\frac{\partial}
{\partial\varphi}\,j_\theta^\varphi(\varphi) = 0\quad ,
\end{equation}
with the current
\begin{equation}
j_\theta^\varphi(\varphi)=\frac{\hbar}{m a}\left(-i\frac
{\partial}{\partial\varphi} - \frac{\phi}{\phi_0} - \frac{1}
{2\cos\theta}\,\sigma_z\right)\sigma_z \quad .
\end{equation}
\end{subequations}
The current of magnetization parallel to the quantization axis in
the local spin frame carried by eigenstates is therefore given by
\begin{equation}
I_{\theta}^{(q\sigma)}=-\frac{1}{e}\frac{\partial E_{q,\sigma}}
{\partial\Phi}\,\sigma\quad .
\end{equation}
Comparison with results from above yield the relation $I_{z}^{(q
\sigma)}=I_{\theta}^{(q\sigma)}\cos\theta$, and we have derived
also the related one $I_{r}^{(q\sigma)}=I_{\theta}^{(q\sigma)}\sin
\theta$.

\begin{figure}[t]
\includegraphics[width=3.in]{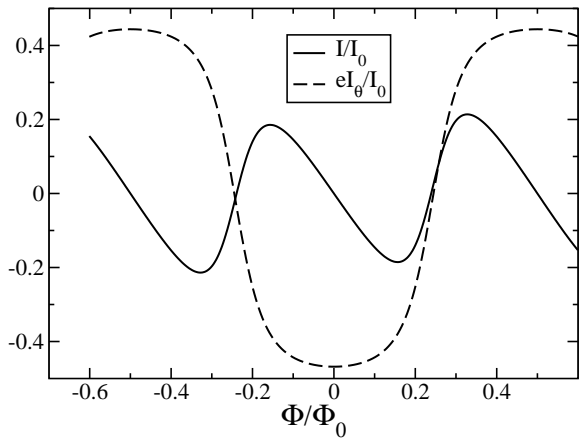}
\caption{Persistent spin current for spin projection onto the
local spin frame (dashed curve) and persistent charge current
(solid curve) vs.\ magnetic flux for the case with electron number
$4N+2$. The barrier strength is $A=0.5$, and $\cos\theta=0.66$.
The current is measured in units of $I_0=\hbar \omega_{\text{a}}
\mathcal{N}/\Phi_0$.
\label{spincurrent}}
\end{figure}  
\begin{figure}[b]
\includegraphics[width=3.in]{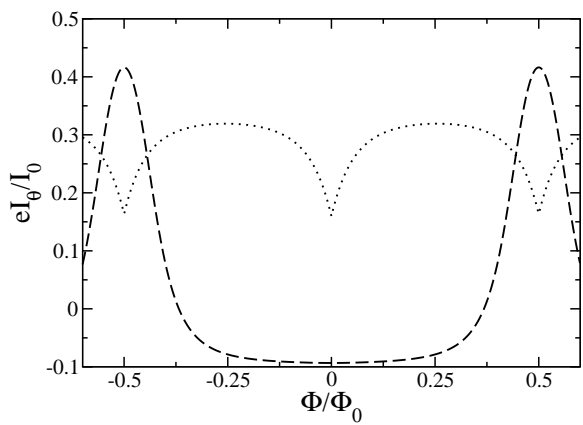}
\caption{Comparison of persistent spin currents for electron
number equal to $4N+2$ (dashed curve) and $2N+1$ (dotted curve).
The barrier strength is $A=0.5$, and $\cos\theta=0.9$
corresponding to a small spin-orbit coupling strength. The
magnitude of persistent spin current decreases rapidly for odd
electron number as $\cos\theta$ approaches $0.66$.
\label{spincompare}}
\end{figure}  

We now present results for the total persistent spin current
$I_\theta=\sum_{q\sigma}I_\theta^{(q\sigma)}$ for the projection
onto the quantization axis of the local spin frame. As shown
above, spin currents for certain other projections can be easily
obtained from $I_\theta$. The fact that flux dependences for the
persistent-current contributions from opposite-spin eigenstates
are shifted according to Eq.~(\ref{flux}) results in large spin
currents at certain flux values. In particular, this is realized
when the currents carried by electrons with opposite spin flow 
in opposite directions. In Fig.~\ref{spincurrent}, we show the
persistent spin current for an even number of electrons. For
comparison, the persistent charge current is plotted as well.
Both exhibit strikingly different flux dependences. Note also
that, in the absence of SO coupling, the persistent spin current
vanishes for even electron number in the ring. Only the relative
shift of energy bands in flux direction caused by SO coupling
enables a finite persistent spin current in this case. For an odd
number of electrons, the persistent spin current is finite both
with and without SO coupling present. We find it to be sizable,
however, only for small values of SO coupling strength. We show a
comparison of even and odd electron number cases in
Fig.~\ref{spincompare}.

The persistent spin current would be a mere theoretical curiosity 
if no detectable effect of it could be found. Fortunately, this is
not so. Recently, it has been pointed out by several
authors\cite{meier-loss,schuetz,hongguo,flavio} that a spin
current, being a magnetization current, gives rise to an electric
field. This is easily proved by making a Lorenz transform to the
rest frame of spin. For example, the electrostatic potential for a
point at a distance $z\ll a$ from the plane of the ring on the
vertical from the center of the ring is 
\begin{equation}\label{elecfield}
\phi(z)\approx\frac{\mu_0}{4\pi}\, g \mu_{\text{B}}\, I_{\theta}\,
\sin \theta \, \frac{a} {z^2}, 
\end{equation}
where $\mu_0$ is the vacuum permeability, $g$ the gyromagnetic
ratio, $\mu_{\text{B}}$ the Bohr magneton, $a$ the radius of the
ring, and $\theta$ the tilt angle due to SO coupling. This result
is identical with the one derived in Ref.~\onlinecite{schuetz} for
the electric field resulting from persistent spin currents in
Heisenberg rings.

\section{Effect of many radial subbands}
\label{sec2d}

\begin{figure}[b]
\includegraphics[width=3.in]{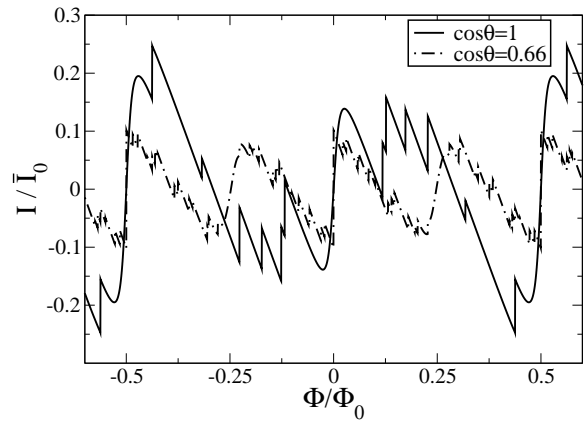}
\caption{Average persistent current vs.\ magnetic flux for a ring
with two occupied radial subbands. The barrier strength is $A=0.1
$. The average is performed on an ensemble containing rings with
occupancy ranging from 60 to 80 electrons. 
\label{twoband}}
\end{figure}  
In the previous section, we have analyzed the persistent current
in a strictly 1D ring, i.e., a ring with only the lowest radial 
subband occupied by electrons and a sufficiently large
subband-energy splitting. We now generalize this discussion to the
case where higher subbands are important. SO coupling introduces
coupling between neighboring radial subbands as described in
Eq.~(\ref{hmix}). More specifically, the Hamiltonian
Eq.~(\ref{hmix}) couples radial subbands with opposite spin 
in the local spin frame, leading to non-parabolicity of energy 
dispersions and to hybridization of opposite-spin bands. The
physics in the limit of strong subband coupling is analogous to 
what happens in a quantum wire with Rashba SO coupling; this
has been discussed in Refs.~\onlinecite{rashbawire,mireles}. 
Here it is sufficient to notice that $H_{n,n+1}$ is negligible if
$l_\omega/l_{so}\ll 1$, i.e., if the radial width of the wave
function is much smaller than the spin-precession length. This
condition is fulfilled in realistic samples. Therefore, we neglect
in the following  the coupling term Eq.~(\ref{hmix}). For the sake
of simplicity we now consider only the two lowest subbands.
Furthermore  we introduce a barrier in the same way as in Section
\ref{1dspectrum}. Assuming that the barrier does not couple
different subbands, and that the transmission coefficient is the  
same for both radial subbands and is given by $t=[1-i \text{sign}
(\kappa)A]/(A^2+1)$, we find for the energy spectrum
\begin{equation}
\label{enkappa2}
E_{q,\pm,n}=\hbar \omega_{\text{a}} \left[\kappa_{q,\pm}^2+\frac
{1}{4}(1-\frac{1}{\cos^2\theta})\right]+\hbar\omega\left(n+\frac
{1}{2}\right), 
\end{equation}
where $n=0,1$ is the subband index, and $\kappa_{q,\pm}$ is still
given by  Eq.~(\ref{solution}). In Fig.~\ref{twoband}, we show the
average persistent current with and without SO coupling. In
comparison to the single-subband case, additional fine structure
appears due to crossing of levels with different radial quantum
numbers. The jumps arising from these extra crossings are very
sharp due to the way we model the barrier, and occur at flux
values that are strongly dependent on the ring occupancy. All 
other features discussed for the strictly 1D case occur at the
same flux values for all radial subbands. Hence, upon averaging,
the latter are magnified and the former demagnified, as it is 
evident comparing  Fig.~\ref{twoband} with Fig.~\ref{average}~a. 
The dependence of the average persistent current on the SO
coupling and barrier strength is the same as for the 1D case,
hence, we do not show it again for the many-subband case. The
presence of many radial subbands, although it introduces some
additional fine structure, essentially yields, after averaging
over different electron numbers, the same SO-related features
discussed in the purely 1D case.

\section{Conclusions} 
\label{conclusions}

We have investigated the effect of Rashba spin-orbit coupling on
the persistent spin and charge currents circling in ballistic
quasi-onedimensional rings. The flux dependence of persistent
charge currents exhibits features that allow for a direct
measurement of the spin-orbit coupling strength. These features
survive averaging over different electron-number configurations
as well as the inclusion of higher subbands. The most striking
effect of spin-orbit coupling discussed here is the occurrence of
finite persistent {\em spin\/} currents for even electron numbers.
We have carefully derived the correct general form of spin
currents in the presence of spin-orbit coupling. The possibility
to measure persistent {\em spin\/} currents via the electric field
generated by their transported magnetization should make it
possible to unambiguously verify the presence and magnitude of
spin-orbit coupling, namely by the different flux dependences of
persistent spin and charge currents.

\begin{acknowledgments}

We have benefited from useful discussions with D.~Frustaglia,
F.~Meijer, A.~Morpurgo, G.~Sch\"on, and A.~Zaikin. This work was
supported by the DFG Center for Functional Nanostructures at the
University of Karls\-ruhe.

\end{acknowledgments}


\begin{thebibliography}{99}
\bibitem{lossnote}
In a related model of electrons on a ring subject to a textured
external magnetic field, additional constant terms contributing to
the spin current were reported\cite{loss:prl:90}. For the case of
electrons on a ring with Rashba SO coupling which is considered
here, no such terms appear.
\end{thebibliography}
\end{document}